\documentclass[11pt,a4paper]{article}
\usepackage{amsfonts}
\usepackage{amsmath,amssymb}
\usepackage{epsfig,graphicx}

\input{tcilatex}
\begin{document}

\begin{center}
{\huge Gauge Fields as Constrained Composite Bosons }

{\Huge \bigskip \bigskip \bigskip }

\textbf{J.L.~Chkareuli}$^{1,2}$

$^{1}$\textit{Institute of Theoretical Physics, Ilia State University, 0162
Tbilisi, Georgia\ \vspace{0pt}\\[0pt]
}

$^{2}$\textit{Andronikashvili} \textit{Institute of Physics, Tbilisi State
University, 0177 Tbilisi, Georgia\ }

\textit{\bigskip }

\bigskip

\bigskip

\bigskip

\bigskip \bigskip

\bigskip

\bigskip

\bigskip

\textbf{Abstract}

\bigskip

\bigskip
\end{center}

We reconsider a scenario in which photons and other gauge fields appear as
the composite vector bosons made of the fermion pairs that may happen with
or without spontaneous violation of Lorentz invariance. The class of
composite models for emergent gauge fields is proposed, where these fields
are required to be restricted by the nonlinear covariant constraint of type $%
A_{\mu }A^{\mu }=M^{2}$. Such a constraint may only appear if the
corresponding fermion currents in the prototype model, being invariant under
some global internal symmetry $G$, are properly constrained as well. In
contrast to the conventional approach, the composite bosons emerged in this
way appear naturally massless, the global symmetry $G$ in the model turns
into the local symmetry $G_{loc}$, while the vector field constraint reveals
itself as the gauge fixing condition. Finally, we consider the case when the
constituent fermions generating emergent gauge bosons could be at the same
time the preons composing the known quark-lepton species in the Standard
Model and Grand Unified Theories.

\bigskip

\thispagestyle{empty}\newpage

\bigskip

%
\thispagestyle{empty}

\section{Introduction}

One can think that local gauge invariance, contrary to a global symmetry
case, may look like a cumbersome geometrical input rather than a true
physical principle, especially in the framework of an effective quantum
field theory becoming, presumably, irrelevant at very high energies. In this
connection, one could wonder whether there is any basic dynamical reason
that necessitates gauge invariance and an associated masslessness of gauge
fields as some emergent phenomenon arising from a more profound level of
dynamics related to their truly elementary constituents. By analogy with a
dynamical origin of massless scalar particle excitations, which is very well
understood in terms of spontaneously broken global internal symmetries \cite%
{NJL}, one could think that the origin of composite massless gauge fields as
the vector Nambu-Goldstone (NG) bosons are presumably related to spontaneous
violation of Lorentz invariance which is in fact the minimal spacetime
global symmetry underlying particle physics. This well-known approach which
might in principle provide a viable alternative to quantum electrodynamics 
\cite{bjorken}, gravity \cite{ph} and Yang-Mills theories \cite{eg,suz} has
a long history started over fifty years ago.

This scenario proposes that a given current-current interaction of basic
constituent fermions induces some almost gauge invariant effective theory
which, apart from the invariant kinetic terms, contain the nontrivial
vector-field potential terms. These terms, no matter whether they exist at
the tree level or appear through radiative corrections, may lead in
principle to spontaneous Lorentz invariance violation (SLIV) according to
which some of components of the emergent composite vector bosons can be
viewed as the corresponding NG zero modes. It is worth pointing out that one
can generally talk about SLIV regardless of whether it is observable or not.
In both of cases, the corresponding zero modes associated with gauge fields
are necessarily generated. Indeed, in the gauge invariance limit this
violation may normally be hidden in gauge degrees of freedom of emergent
vector fields. However, when these superfluous degrees are eliminated by
gauge symmetry breaking, SLIV becomes observable. In this connection, we
will use later on the terms \textquotedblleft inactive
SLIV\textquotedblright\ and \textquotedblleft active SLIV\textquotedblright
, respectively, in order to distinguish these two cases.

The important point, however, is that the potential-based vector field
models which can only lead to the active or physical SLIV appear to be
generically unstable and contradictory and, therefore, seems to be hardly
acceptable in their present form, as we will argue below. On the other hand,
why should one necessarily insist on physical Lorentz violation, if emergent
gauge fields are anyway generated through the \textquotedblleft
safe\textquotedblright\ inactive SLIV models which recover a conventional
Lorentz invariance?

In the light of this reasoning, we propose here an alternative approach for
composite gauge bosons replacing the potential-based vector field models by
the constraint-based ones which we briefly sketch below in Sections 2 and 3,
respectively. Later in Section 4, we show how such constrained gauge bosons
are induced by the properly constrained currents of the constituent fermions
involved. In contrast to the potential-based approach, the composite gauge
bosons appear now to be naturally massless, though being restricted by some
nonlinear constraint which is then treated as their nonlinear gauge. So, the
starting global symmetry of constituent fermions $G$ turns into the local
internal symmetry $G_{loc}$ of the gauge and matter fields involved.
Further, in Section 5, we discuss a possible scenario when the constituent
fermions generating emergent gauge bosons could be at the same time the
preons composing the known quark-lepton species in the Standard Model and
Grand Unified Theories. And, finally, our conclusion is provided in Section
6, where we discuss why basically the proposed constraint pattern for vector
fields could make the whole physical field system involved to adjust itself
in a gauge invariant way. Otherwise, as is argued, it could loose too many
degrees of freedom, thus getting unphysical.

\section{Models with vector field potential}

For an emergence of the potential-based vector field models one can start
with the generalized prototype Lagrangian with all possible multi-fermi
current-current interactions \cite{kraus} rather than with the original
four-fermi one \cite{bjorken} 
\begin{equation}
L(\psi )=\text{ }\overline{\psi }_{s}(i\gamma \partial -m)\psi
_{s}+N\sum_{p=1}^{\infty }\mathrm{G}_{p}\left( j_{\mu }j^{\mu }/N^{2}\right)
^{p}\text{ }  \label{4f}
\end{equation}%
where the fermion set includes $N$ constituent fermion species $\psi _{s}$
(hereafter, summation over all repeated indices is implied). The Lagrangian $%
L(\psi )$ possesses the $U(N)$ global flavor symmetry under which the above
all-fermion current $j_{\mu }$ 
\begin{equation}
\text{\ }j_{\mu }(\psi )=\overline{\psi }_{s}\gamma _{\mu }\psi _{s}\text{ \
\ \ \ \ }(s=1,2,...,N)  \label{j}
\end{equation}%
is invariant in itself. This model is evidently non-renormalizable and can
only be considered as an effective theory valid at sufficiently low
energies. The dimensionful couplings $\mathrm{G}_{p}$ are proportional to
appropriate powers of the UV cutoff $\Lambda $ being ultimately related to
some energy scale up to which this effective theory is valid, $\mathrm{G}%
_{p}\sim \Lambda ^{4-6p}$. Factors of $N$ in (\ref{4f}) are chosen in such a
way to provide a well defined large $N$ limit.

The Lagrangian (\ref{4f}) can be re-written using the standard trick of
introducing an auxiliary field $A_{\mu }$ 
\begin{equation}
L(\psi ,A_{\mu })=\text{ }\overline{\psi }_{s}(i\gamma \partial -\gamma
A-m)\psi _{s}-N\mathrm{V}(A_{\mu }A^{\mu })  \label{4f1}
\end{equation}%
The potential $V$ is a power series in $A_{\mu }A^{\mu }$ that can generally
be written in the form%
\begin{equation}
\mathrm{V}(A_{\mu }A^{\mu })=\sum_{p=1}^{\infty }\lambda _{p}\left( A_{\mu
}A^{\mu }-M^{2}\right) ^{2p}  \label{exp}
\end{equation}%
with the coefficients $\lambda _{p}$ and mass parameter $M$ chosen such that
by solving the algebraic equations of motion for $A_{\mu }$ and substituting
back into (\ref{4f1}) one recovers the starting Lagrangian (\ref{4f}). If
instead one integrates out the fermions $\psi _{s}$, one gets an effective
action in terms of the properly renormalized composite $A_{\mu }$\ field
which acquires its own dynamics%
\begin{equation}
S_{eff}=N\int d^{4}x\left[ -\frac{1}{4}F_{\mu \nu }F^{\mu \nu }+\mathrm{V}%
(A_{\mu }A^{\mu })+A_{\mu }J^{\mu }\right]  \label{efff}
\end{equation}%
Since the fermions $\psi _{s}$ are minimally coupled to the vector field $%
A_{\mu }$ in (\ref{4f1}), its kinetic term generated in this way appears
gauge invariant provided that a gauge invariant cutoff is chosen.
Furthermore, since there are $N$ species of fermions $\psi _{s}$ the
effective action (\ref{efff}) has an overall factor of $N$. And the last
point is that one has to introduce in the basic Lagrangian (\ref{4f}), apart
from the pure constituent fermions $\psi _{s}$, some actual matter fields $%
\Psi _{t}$ interacting through their own conserved current $J_{\mu }$ 
\begin{equation}
\text{\ }J_{\mu }(\Psi )=\overline{\Psi }_{t}\gamma _{\mu }\Psi _{t}\text{\
\ }(t=1,2,...,N_{\Psi })  \label{jj}
\end{equation}%
This will generate in turn the standard gauge invariant matter coupling
given in the action (\ref{efff}). Indeed, in contrast to the heavy fermions $%
\psi _{s}$ whose masses are normally proposed to be of the cutoff order, the
matter fermions $\Psi _{t}$ can be taken to be light and even massless. This
allows to keep them in the low energy effective action, whereas the heavy
fermions $\psi _{s}$ are integrated out. They are in some sense hidden ones
which are solely needed to properly induce the composite vector bosons as
appropriate gauge fields. Note that, while their number $N$ is in fact
somewhat technical one allowing to properly suppress the high-order
dangerous terms, $N_{\Psi }$ is a number of the actually observed similar
matter fermion species, say, those being among known quarks and leptons.

At the first glance, everything looks good in the effective action (\ref%
{efff}). However, some generic problems related to SLIV\ and stability of
the emerged theory may necessarily appear. Let us consider at the beginning
only the first non-trivial term in the multi-fermi interactions (\ref{4f})
and, respectively, in the potential $\mathrm{V}$ (\ref{efff}) that just
corresponds to the original Bjorken model \cite{bjorken}. As matter of fact,
this model naturally leads to the massive QED theory rather than to the
conventional massless QED one. Indeed, any conclusion that such a massive
vector boson might be condensed due to its radiatively produced quartic term 
\cite{bjorken} or even be massless in itself through a strict cancellation
of its tree-level and radiative masses \cite{eg}, seems to be rather
problematic since it is related to somewhat formal manipulations with
divergent integrals. The same could be said about the non-Abelian symmetry
case as well\footnote{%
An interesting non-Abelian model was presented in \cite{suz}. It starts with
current-current interaction involving some large $N$ sets of fermions
assigned to the fundamental representation of some $SU(n)$ group. It was
then shown that in the leading $N$ order, an explicit computation of the
infinite fermion chain allows to completely reproduce the massive $SU(n)$
Yang-Mills theory. The composite boson mass can not be made zero for any
finite value of the binding current-current coupling constant in the
starting fermion Lagrangian. Thus, as it happens, this emergent theory does
not possess a true gauge symmetry as well.}.

It might seem that the above problem would be over when the higher-order
terms beyond the four-fermi interaction are activated in the basic fermion
Lagrangian (\ref{4f}) and, respectively, in the potential (\ref{exp}). As is
readily seen from (\ref{exp}), the next term in it really gives the quartic $%
A_{\mu }$ field term in the effective action $S_{eff}$ (\ref{efff}). This is
enough to generate the familiar Mexican hat structure of the potential 
\textrm{V}$(A_{\mu }A^{\mu })$, thus coming to spontaneous Lorentz violation
at a scale determined by the mass parameter $\left\vert M\right\vert $.
Rewriting the action (\ref{efff}) in terms of the renormalized $A_{\mu }$
field and leaving in the potential only bilinear and quartic vector field
terms one gets the effective theory being sometimes referred to as the
\textquotedblleft bumblebee\textquotedblright\ model \cite{ks}. This
partially gauge invariant model means in fact that the vector field $A_{\mu
} $ develops a constant background value and Lorentz symmetry $SO(1,3)$
breaks down to $SO(3)$ or $SO(1,2)$ depending on whether the $M^{2}$ is
positive or negative, respectively. In both of cases, there are three zero
massless modes and a heavy Higgs mode in the symmetry broken phase.

The point is, however, that not only this bumblebee-like model but all
possible potential-based models appear generally unstable. Their
Hamiltonians (as was argued specifically for the bumblebee case \cite{vru}
but those arguments are applicable to a general $\mathrm{V}$ potential as
well) are not bounded from below beyond the constrained phase space
determined by the nonlinear condition put on the vector field, 
\begin{equation}
A_{\mu }A^{\mu }=M^{2}  \label{cd}
\end{equation}%
as can be readily shown using Dirac constraint analysis \cite{dr}. With this
condition imposed, the massive Higgs mode never appears, the Hamiltonian
turns to be positive, and the model is physically equivalent to the
nonlinear constraint-based QED, which we consider in the next section.

Note also that, apart from the instability, these models, in contrast to
some viable effective field theories, do not posses a consistent ultraviolet
completion \cite{aki}. This makes their application rather problematic in a
sense that one can not draw relevant conclusions about the strength of any
physical effects in such an effective theory.

And one more argument against the potential-based models may follow from
their supersymmetric extension. As one can readily confirm, SUSY may only
admit the bilinear mass term in the vector field potential energy (see \cite%
{c} for more details). As a result, without the stabilizing quartic (and
higher order) vector field terms, this type of spontaneous Lorentz violation
can in no way be realized in the SUSY context. The same could be said about
the prototype Lagrangian with the multi-fermi current-current interactions (%
\ref{4f1}) which can not be constructed from any matter chiral superfields.

All that means that the above-mentioned composite models leading eventually
to the vector field potential $\mathrm{V}(A_{\mu }A^{\mu })$, whether it
contains only the mass term or the higher order terms as well, seems to be
hardly acceptable in the present form. One might only expect that, since a\
mass scale $\left\vert M\right\vert $ of SLIV is normally presumed to be
near the Planck scale, the quantum gravity theory would make the ultimate
conclusion on physical viability of such models.

\section{Models with vector field constraint}

We give here some brief sketch of the constraint-based models for vector
fields considering first the constraint-based QED model. This model starts
directly through the vector field \textquotedblleft
length-fixing\textquotedblright\ constraint (\ref{cd}) implemented into the
conventional QED. Such type of models were first studied by Dirac \cite{dir}
and Nambu \cite{nambu} a long ago, and in more detail in recent years \cite%
{az,ks1,jej,cj,wiki,gra,urr,esc}.

The constraint (\ref{cd}) is in fact very similar to the constraint
appearing in the nonlinear $\sigma $-model for pions, $\sigma ^{2}+\pi
^{2}=f_{\pi }^{2}$, where $f_{\pi }$ is the pion decay constant\cite{GL}. As
is well know, this constraint leads to spontaneous breaking of the
underlying chiral symmetry $SU(2)\times SU(2)$ in the model. Analogously, as
Nambu argued \cite{nambu}, the constraint (\ref{cd}) might lead to
spontaneous Lorentz violation. Rather than impose by postulate, the
constraint (\ref{cd}) may be implemented through the Lagrange multiplier
term into the standard QED Lagrangian

\begin{equation}
L(\Psi ,A,\mathrm{\lambda })=-\frac{1}{4}F_{\mu \nu }F^{\mu \nu }+\overline{%
\Psi }_{t}(i\gamma \partial -e\gamma A-\mathrm{m})\Psi _{t}-\frac{\mathrm{%
\lambda }}{2}\left( A_{\mu }A^{\mu }-M^{2}\right) \text{ }  \label{lag}
\end{equation}%
for some charged matter fermion fields $\Psi _{t}$ ($t=1,2,...,N_{\Psi }$).
The variation under the multiplier field $\mathrm{\lambda }(x)$ leads then
to the vector field constraint (\ref{cd}).

Actually, this Lagrangian, when taken without matter, is the original Dirac
theory \cite{dir} proposed for an alternative introduction of classical
electric charge. In fact, the Lagrange multiplier term in (\ref{lag})
corresponds to interaction of vector field with the extra source current $%
J_{\mu }^{ext}=\mathrm{\lambda }A_{\mu }$. However, when the charged matter
fields are included there appears the vector field interaction with the
conventional Noether current $J_{\mu }=\overline{\Psi }_{t}\gamma _{\mu
}\Psi _{t}$ as well. As follows from the total Lagrangian (\ref{lag}), both
of them are separately conserved 
\begin{equation}
\partial ^{\mu }J_{\mu }^{ext}=0\text{ , \ \ }\partial ^{\mu }J_{\mu }=0%
\text{\ }  \label{ej}
\end{equation}%
This extended Dirac theory contains five equations for the five field
quantities, $A_{\mu }$ and $\mathrm{\lambda }$, that includes their
equations of motion and the extra current conservation (\ref{ej}). The
solutions of these equations are fixed when the appropriate initial
conditions are prescribed. We propose, as in the original Dirac model \cite%
{dir}, that the starting values for all fields (and their momenta) involved
are chosen so as to restrict their phase space to values providing the
infinitesimal multiplier function \textrm{$\lambda $}$(x)$. Remarkably, due
to an automatic conservation of the extra source current, $J_{\mu }^{ext}$ (%
\ref{ej}), 
\begin{equation}
\partial \lambda /\partial t=[\partial _{i}(\lambda A_{i})-\lambda (\partial
_{0}A_{0})]/A_{0}  \label{ej1}
\end{equation}%
such type of $\mathrm{\lambda }$ field will then remain for all times. One
can thus safely keep the multiplier term in the Lagrangian (\ref{lag}) to
derive the vector field constraint (\ref{cd}) in an ordinary way. This
Lagrangian, being the ground for our further consideration, goes to the
standard QED in the vanishing $\mathrm{\lambda }$ limit being provided again
by the choice of the proper initial conditions. The only difference is that
the vector potentials must now satisfy not only their equations of motion
but the constraint (\ref{cd}) as well\footnote{%
Note that an arbitrary finite $\mathrm{\lambda }$ field may generally cause
instability of the theory making its Hamiltonian negative \cite{vru}.
However, with the infinitesimal \textrm{$\lambda $ }field chosen (that is
quite enough to provide the vector field constraint (\ref{cd})) the
Hamiltonian is always positive once Gauss' law holds \cite{dir}. It is also
worth pointing out that even the zero limit for the infinitesimal \textrm{$%
\lambda $ }field is quite imaginable, since in this limit the above extra
charges are successively disappeared, while the vector field constraint
remains \cite{dir}.}.

One way or another, the constraint (\ref{cd}) means in essence that the
vector field $A_{\mu }$ develops presumably the VEV along the direction
given by the unit Lorentz vector $n_{\mu }$%
\begin{equation}
\left\langle A_{\mu }\right\rangle =n_{\mu }M\text{ \ \ \ \ \ }(n^{2}=n_{\mu
}n^{\mu }=1)  \label{vev1}
\end{equation}%
causing for certainty some time-like ($M^{2}>0$) Lorentz violation at a
scale $M$, while rotational invariance is still maintained. This type of
SLIV produces an ordinary photon as a true Goldstone vector boson ($a_{\mu }$%
) being orthogonal to the vacuum direction given by the vector $n_{\mu }$ 
\begin{equation}
A_{\mu }=a_{\mu }+n_{\mu }\sqrt{M^{2}-a^{2}}\text{ , \ }n_{\mu }a_{\mu }=0%
\text{ \ }(a^{2}\equiv a_{\mu }a^{\mu })  \label{gol}
\end{equation}%
The point is, however, that in sharp contrast to the nonlinear $\sigma $
model for pions, the constrained QED theory (\ref{lag}) ensures that
physical Lorentz invariance in it remains unbroken. Indeed, although the
theory in the symmetry broken phase contains a plethora of Lorentz and $CPT$
violating couplings when it is expressed in terms of emergent $a_{\mu }$
modes, the contributions of all these Lorentz violating couplings to
physical processes completely cancel out among themselves. Actually, as was
shown in the tree \cite{nambu} and one-loop approximations \cite{az}, the
nonlinear constraint (\ref{cd}) applied as a supplementary condition appears
in essence as a possible gauge choice for the vector field $A_{\mu }$, while
the $S$-matrix remains unaltered under such a gauge convention. The similar
result was also confirmed for spontaneously broken non-Abelian theories \cite%
{jej,cj} and tensor field gravity \cite{gra}.

Let us describe in some detail just the\ non-Abelian symmetry case since we
are going to consider later the Yang-Mills theories with composite gauge
fields. Let us assume there is such a theory possessing some internal
symmetry $G$ having $D$ generators so that the nonlinear constraint for the
vector field multiplet has now the form 
\begin{equation}
A_{\mu }^{i}A^{\mu i}=M^{2}\text{ \ \ (}i=1,2,...,D\text{)}  \label{aa}
\end{equation}%
Remarkably, this time not only the pure Lorentz symmetry $SO(1,3)$, but the
much larger accidental symmetry $SO(D,3D)$ of the SLIV constraint (\ref{aa})
also happens to be spontaneously broken. As a result, although the pure
Lorentz violation still generates only one true Goldstone vector boson, the
accompanying pseudo-Goldstone vector bosons related to the $SO(D,3D)$
breaking 
\begin{equation}
A_{\mu }^{i}=a_{\mu }^{i}+n_{\mu }^{i}\sqrt{M^{2}-a^{2}}\text{ , \ }n_{\mu
}^{i}a_{\mu }^{i}=0\text{ \ }(n^{2}\equiv n_{\mu }^{i}n^{i\mu }=1\text{, }%
a^{2}\equiv a_{\mu }^{i}a^{i\mu })  \label{gol1}
\end{equation}%
also come into play properly completing the whole gauge multiplet of the
internal symmetry group $G$ taken. In contrast to the known scalar
pseudo-Goldstone modes, they remain strictly massless, being protected by
the simultaneously generated non-Abelian gauge invariance \cite{jej,cj}.

To conclude, the constraint-based theories are in fact emergent gauge
theories appearing due to spontaneous Lorentz violation which we refer to as
inactive SLIV. These theories, both Abelian and non-Abelian, when being
expressed in terms of the pure Goldstone vector modes ($a_{\mu }$ and $%
a_{\mu }^{i}$, respectively) look essentially nonlinear and contain in
general a variety of Lorentz and $CPT$ breaking couplings. However, due to
total cancellations of their contributions among themselves, they appear to
be physically indistinguishable from the conventional QED and Yang-Mills
theories. Their emergent nature could only be seen when taking the covariant
gauge condition (\ref{cd}) into account. Any other gauge, e.g. Coulomb
gauge, is not in line with emergent picture, since it \textquotedblleft
breaks\textquotedblright\ Lorentz invariance in an explicit rather than
spontaneous way. As to an observational evidence in favor of emergent
theories, the only way for SLIV to become active, thus causing physical
Lorentz violation, would only appear if gauge invariance in these theories
were really broken \cite{par} rather than merely constrained by the
gauge-fixing term. In substance, the above SLIV ansatz, due to which the
vector field $A_{\mu }(x)$ develops the VEV (\ref{vev1}), may itself be
treated as a pure gauge transformation with a gauge function linear in
coordinates, $\omega (x)=$ $n_{\mu }x^{\mu }M$. From this viewpoint, gauge
invariance in QED or Yang-Mills theory leads to the conversion of SLIV into
gauge degrees of freedom of the massless gauge fields emerged. This is what
one could refer to as the generic non-observability of SLIV in gauge
invariant theories. Moreover, as was shown some time ago \cite{cfn}, gauge
theories, both Abelian and non-Abelian, can be obtained by themselves from
the requirement of the physical non-observability of SLIV induced by
condensation of vector fields rather than from the standard gauge principle.

\section{Constrained composite bosons}

We are coming now to the most interesting question: how can the constrained
QED Lagrangian (\ref{lag}) be generated from some underlying dynamics of the
elementary constituent fermions? One can readily confirm following the
standard procedure that such basic Lagrangian for both hidden constituent
and real matter fermions, $\psi _{s}$ and $\Psi _{t}$, has to have the form 
\begin{equation}
L(\psi ,\Psi ,\mathrm{\lambda })=\bar{\psi}_{s}(i\gamma \partial -m)\psi
_{s}+\overline{\Psi }_{t}(i\gamma \partial -\mathrm{m})\Psi _{t}+\frac{1}{2%
\mathrm{\lambda }}(j_{\mu }+J_{\mu })^{2}+\frac{\mathrm{\lambda }}{2}M^{2}%
\text{ }  \label{fF}
\end{equation}%
where $j_{\mu }$ and $J_{\mu }$ stand for their Noether currents (\ref{j})
and (\ref{jj}), respectively. Indeed, employing the standard trick of
introducing an auxiliary field $A_{\mu }(x)$ one has instead%
\begin{equation}
L(\psi ,\Psi ,\mathrm{\lambda ,}A_{\mu })=L_{0}(\psi )+L_{0}\left( \Psi
\right) -(j_{\mu }+J_{\mu })A^{\mu }-\frac{\mathrm{\lambda }}{2}(A_{\mu
}A^{\mu }-M^{2})  \label{ff1}
\end{equation}%
with $L_{0}(\psi )$ and $L_{0}\left( \Psi \right) $ being the free
Lagrangians for fermions $\psi _{s}$ and $\Psi _{t}$. Now, solving the
algebraic equations of motion for $A^{\mu }$, 
\begin{equation}
A_{\mu }=-\frac{1}{\mathrm{\lambda }}(j_{\mu }+J_{\mu })  \label{f3}
\end{equation}%
and substituting back into the Lagrangian (\ref{ff1}) one recovers the
starting fermion Lagrangian (\ref{fF}). Remarkably, as one can readily see,
the equation (\ref{f3}) connects the Dirac extra source current (\ref{ej})
with the standard fermion currents in the theory%
\begin{equation}
j_{\mu }+J_{\mu }=-\text{ }J_{\mu }^{ext}  \label{rel}
\end{equation}%
Now, variation of the prototype vector field Lagrangian (\ref{ff1}) under
the multiplier function $\mathrm{\lambda }$ leads to the vector field
constraint (\ref{cd}), mentioned above, while a similar variation of the
fermion Lagrangian (\ref{fF}) gives in turn the constraint for the total
fermion current in itself%
\begin{equation}
(j_{\mu }+J_{\mu })^{2}=\mathrm{\lambda }^{2}M^{2}  \label{cj}
\end{equation}

And eventually, integrating all constituent fermions $\psi _{s}$ out from
the prototype Lagrangian (\ref{ff1}) we come to the effective Lagrangian (%
\ref{lag}) expressed in terms of the properly renormalized composite $A^{\mu
}$\ field interacting with matter fermions $\Psi _{t}$ in gauge invariant
way. Indeed, since the constituent fermions $\psi _{s}$ were minimally
coupled to $A_{\mu }$ in the prototype Lagrangian, all the generated terms
in the effective Lagrangian (including any high-order ones) appear gauge
invariant provided a gauge invariant cutoff is chosen. The vector field
kinetic term appears in the form 
\begin{equation}
-Z_{3}\times \frac{1}{4}F_{\mu \nu }F^{\mu \nu }\text{, \ }Z_{3}=\frac{N}{%
12\pi ^{2}}\ln \frac{\Lambda ^{2}}{m^{2}}  \label{r}
\end{equation}%
where the renormalization constant $Z_{3}$ is given by the usual vacuum
polarization integral with some momentum-space cutoff $\Lambda $. The $Z_{3}$
is then absorbed into the wave-function renormalization of $A^{\mu }$\ field
which in turn renormalizes the gauge coupling, multiplier field and SLIV
scale as well%
\begin{equation}
A_{\mu }\rightarrow A_{\mu }/\sqrt{Z_{3}}\text{ , \ }1\rightarrow e=1/\sqrt{%
Z_{3}}\text{ , \ }\mathrm{\lambda }\rightarrow \mathrm{\lambda }/Z_{3}\text{
, \ }M^{2}\rightarrow M^{2}Z_{3}  \label{rr}
\end{equation}%
that leads eventually to the gauge invariant effective Lagrangian (\ref{lag}%
) (with all the former notations remained). \ The nonlinear vector field
constraint term in it, though being noninvariant, appears in fact as gauge
condition in an otherwise gauge invariant and Lorentz invariant theory. So,
in sharp contrast to the potential-based models considered above, all
radiative corrections which may appear in the effective Lagrangian (\ref{lag}%
) have to necessarily be both Lorentz invariant and gauge invariant. In
fact, in one-loop approximation it was explicitly demonstrated some time ago 
\cite{az}.

Interestingly, while the nonlinear vector field constraint (\ref{cd}) turns
into the gauge condition, the constraint (\ref{cj}) put on the currents
could mean some relation between currents of the hidden constituents and
matter fermions. Actually, this relation also includes an extra source of
charge density given by the $\mathrm{\lambda }$ field which is taken to be
infinitesimal in our model. This means that the hidden and the actual matter
currents appear to be approximately opposite to each other, $j^{\mu }\approx
-J^{\mu }$. In this connection, one could notice that our basic
field-current identity (\ref{f3}) is in fact the ratio of two infinitesimal
field quantities giving eventually the finite vector field $A_{\mu }$. While
in the prototype Lagrangian (\ref{ff1}) it couples with the infinitesimal
sum of the currents $j_{\mu }+J_{\mu }$, in the effective Lagrangian (\ref%
{lag}) emerging after integrating out of the hidden constituent fermions $%
\psi _{sa}$ the vector field interaction is solely given by the finite
coupling with the matter fermions 
\begin{equation}
A_{\mu }J^{\mu }=A_{\mu }\overline{\Psi }_{t}\gamma ^{\mu }\Psi _{t}\text{\
\ \ }(t=1,2,...,N_{\Psi })  \label{cj1}
\end{equation}

And last but not least, it is worth pointing out that the constraint-based
models possess one more remarkable advantage as compared to the
potential-based ones -- they do not need in general to have the enormously
large number $N$ of the hidden constituent fermion species. Actually such a
number $N$ was mainly imposed to properly suppress the large physical
Lorentz violation at low energies which could otherwise appear through the
uncontrollably large radiative corrections to the effective Lagrangians.
That is not expected at all in the constraint-based models and, therefore,
the number of the constituent fermions $\psi _{sa}$ $(s=1,2,...,N)$\ \ may
be determined now by only their own dynamics.

Let us now briefly describe a possible extension of this approach to the
non-Abelian $G$ symmetry case employing the non-Abelian model with
constrained vector field multiplet that we discussed in the previous
section. Now, all the $N$ and $N_{\Psi }$ species of hidden constituent and
matter fermions, $\psi _{s}$ and $\Psi _{t}$ belong to the fundamental
representations of $G$, $\psi _{sa}$ and $\Psi _{ta}$ ($a=1,2,...,n$), while
the emerging composite vector fields will complete its adjoint multiplet.
This extension can readily be made just by the corresponding replacements in
the above equations (\ref{fF}-\ref{cj})%
\begin{equation}
A_{\mu }\rightarrow A_{\mu }^{i}\text{ , \ }j_{\mu }\rightarrow j_{\mu }^{i}%
\text{ , \ \ }J_{\mu }\rightarrow J_{\mu }^{i}\text{ \ (}i=1,2,...,D\text{)}
\label{aA}
\end{equation}%
($D$ stands for the $G$ symmetry group dimension). One can then proceed in a
conventional way \cite{eg} to generate the vector field kinetic energy term
together with the three- and four-vector self-couplings from the relevant
loop diagrams. We eventually arrive at the totally gauge invariant theory
with an obvious replacement in the QED Lagrangian (\ref{lag}), $F_{\mu \nu
}\rightarrow F_{\mu \nu }^{i}$, where $F_{\mu \nu }^{i}$ is the standard
strength-tensor for the constrained vector field multiplet $A_{\mu }^{i}$.
The modified Lagrange multiplier term providing the constraint (\ref{aa})
appears again as the pure gauge-fixing condition.

\section{Composite quarks, leptons and gauge bosons}

We present here some scenario how one could in principle combine this type
models of composite gauge bosons with models for quarks and leptons composed
from preons (some significant references can be found in \cite{pr}). The
preons are usually viewed as the truly elementary carriers of all known
physical charges, such as weak isospin, color, family number, etc. (which we
refer to as \textquotedblleft metaflavors\textquotedblright ). Apart from
metaflavors, they normally possess some metacolor forces that bind them
inside quarks and leptons. And last but not least, they should be massless
(or having some tiny masses) that is required to have at large distances the
observed quarks and leptons with masses which much less than their
composition scale. For that, as is well known, the chiral symmetry of the
preons should be remained so that the anomaly matching condition of preons
at small distances and their composites at the large ones has to be
satisfied \cite{th}.

In the light of this, the above massive constituent fermions $\psi _{s}$
composing gauge bosons can not be used for composition of quarks and
leptons. The most direct way would be to treat the matter fields as the
massless preon candidates supplying them with both metacolor and metaflavor
symmetries $G_{MC}$ and $G_{MF}$%
\begin{equation}
\Psi _{ak}\text{ , \ }a=1,2,...,n_{MC}\text{ , \ }k=1,2,...,n_{MF}
\label{mf}
\end{equation}%
where indices $a$ and $k$ belong to their fundamental representations,
respectively. They are still global symmetries which then will be converted
into the local ones by the proper current-current interaction of the hidden
constituent fermions, such as it was described above. We propose that they
are assigned to the symmetry groups $G_{MC}$ and $G_{MF}$ in the same way%
\begin{equation}
\psi _{ak}\text{ , \ }a=1,2,...,n_{MC}\text{ , \ \ }k=1,2,...,n_{MF}
\label{cf}
\end{equation}%
Note that in the both cases (\ref{mf}) and (\ref{cf}), we identify the
number of fermions, the hidden constituent and matter ones, with number of
metaflavors in the symmetry group $G_{MF}$%
\begin{equation}
N=N_{\Psi }=n_{MF}  \label{Nn}
\end{equation}%
One might, of course, introduce some independent large $N$ number of the
hidden constituent fermions (\ref{cf}). However, as was mentioned above, in
the constraint-based models, one can have it as low as it is required by the
composite dynamics itself.

So, technically, we can start with the pure fermion Lagrangian like that of (%
\ref{fF}) but being extended to the non-Abelian symmetry corresponding to
the still global metacolor and metaflavor symmetries $G_{MC}$ and $G_{MF}$,
respectively. One can then readily use the above procedure (\ref{ff1}, \ref%
{f3}) of introducing the auxiliary vector fields

\begin{equation}
A_{\mu }^{i}=A_{b\mu }^{a}(T^{i})_{a}^{b}\text{ , \ }\mathcal{A}_{\mu }^{r}=%
\mathcal{A}_{l}^{k}(\mathcal{T}^{r})_{k}^{l}\text{ \ \ }(i=1,2,...,D_{MC}%
\text{ };\text{ \ \ }r=1,2,...,D_{MF})  \label{aaa}
\end{equation}%
in the Lagrangian through the interrelated Noether currents of the above
fermions (\ref{mf}) and (\ref{cf})%
\begin{eqnarray}
A_{\mu }^{i} &=&-\frac{1}{2\mathrm{\lambda }_{MC}}(j_{\mu }^{i}+J_{\mu }^{i})%
\text{ , \ }j_{\mu }^{i}=\overline{\psi }\gamma _{\mu }T^{i}\psi \text{ , \ }%
J_{\mu }^{i}=\overline{\Psi }\gamma _{\mu }T^{i}\Psi  \notag \\
\text{\ }\mathcal{A}_{\mu }^{r} &=&-\frac{1}{2\mathrm{\lambda }_{MF}}(%
\mathrm{j}_{\mu }^{r}+\mathcal{J}_{\mu }^{r})\text{ , \ }\mathrm{j}_{\mu
}^{r}=\overline{\psi }\gamma _{\mu }\mathcal{T}^{r}\psi \text{ , \ }\mathcal{%
J}_{\mu }^{r}=\overline{\Psi }\gamma _{\mu }\mathcal{T}^{r}\Psi  \label{bb}
\end{eqnarray}%
Here $T^{i}$ and $\mathcal{T}^{r}$ stand for generators of the symmetry
groups $G_{MC}$ and $G_{MF}$ having dimensions $D_{MC}$ and $D_{MF}$,
respectively, while $\mathrm{\lambda }_{MC}(x)$ and $\mathrm{\lambda }%
_{MF}(x)$ are the corresponding Lagrange multiplier fields. Note also that
the metacolor currents \ $j_{\mu }^{i}$ and $J_{\mu }^{i}$ are invariant
under the metaflavor symmetry $G_{MF}$, as well as the metaflavor currents \ 
$\mathrm{j}_{\mu }^{r}$ \ and $\mathcal{J}_{\mu }^{r}$\ are invariant under
the metacolor symmetry $G_{MC}$ (the corresponding fermion indices are
properly summed up in them).

Further, a subsequent integration of the massive constituent fermion
multiplets $\psi _{ak}$ out will convert the global symmetries $G_{MC}$ and $%
G_{MF}$ into the local ones with the metacolor and metaflavor gauge fields $%
A_{\mu }^{i}$ and $\mathcal{A}_{\mu }^{r}$ interacting with the presumably
massless preon multiplets $\Psi _{ak}$. After the appropriate
renormalization of all the composite vector fields involved one arrives at
the final preon Lagrangian%
\begin{eqnarray}
L(\Psi ,A,\mathcal{A}) &=&L_{0}\left( \Psi \right) -\frac{1}{4}F_{\mu \nu
}^{i}F^{i\mu \nu }-\frac{1}{4}\mathcal{F}_{\mu \nu }^{r}\mathcal{F}^{r\mu
\nu }-gJ_{\mu }^{i}A^{i\mu }-\mathrm{g}\mathcal{J}_{\mu }^{r}\mathcal{A}%
^{r\mu }  \notag \\
&&-\mathrm{\lambda }_{MC}(A_{\mu }^{i}A^{i\mu }-M_{A}^{2})-\mathrm{\lambda }%
_{MF}(\mathcal{A}_{\mu }^{r}\mathcal{A}^{r\mu }-\mathrm{M}_{\mathcal{A}}^{2})
\label{st}
\end{eqnarray}%
with the vector field constraints included, which contain the corresponding
Lagrange multiplier functions $\mathrm{\lambda }_{MC}(x)$ and $\mathrm{%
\lambda }_{MF}(x)$, as well as the SLIV scales $M_{A}$ and $\mathrm{M}_{%
\mathcal{A}}$, respectively. The equations (\ref{bb}) show that both
metacolor and metaflavor gauge field multiplets, $A_{\mu }^{i}$ and $%
\mathcal{A}_{\mu }^{r}$, apart from the hidden constituent fermions $\psi
_{ak}$ (\ref{cf}), are also consisted of the preons $\Psi _{ak}$ (\ref{mf})
which at the same time compose the observed quarks and lepton through the
metacolor forces. It goes without saying that their composition scale $%
\Lambda _{MC}$ is always less than the cutoff $\Lambda $ of the prototype
fermion theory.

After all that, some whole scenario with composite quarks and leptons
interacting with composite gauge bosons may be properly developed depending
on the particular symmetry groups $G_{MC}$ and $G_{MF}$ taken. One possible
scenario could be realized in the model of composite quarks and leptons that
was recently presented in \cite{su}. We give below some of its key elements:

(1) It is proposed that there are $2K$ elementary massless left-handed and
right-handed preons at small distances, $P_{kL}$ and $Q_{kR}$ ($k=1,\ldots
,K $), which possess a common local symmetry $SU(K)_{MF}$ unifying all known
metaflavors, such as weak isospin, color, family number, etc. The preons,
both $P_{kL}$ and $Q_{kR}$, transform under fundamental representation of $%
SU(K)_{MF}$ and their metaflavor theory has presumably an exact $L$-$R$
symmetry. Actually, the $SU(K)_{MF}$\ appears at the outset as some
vectorlike symmetry which then breaks down at large distances to some of its
chiral subgroup.

(2) In contrast to their common metaflavors, the left-handed and
right-handed preon multiplets are taken to be chiral under the local
metacolor symmetry $G_{MC}=SO(3)_{L}\times SO(3)_{R}$. They appear with
different metacolors, $P_{kL}^{a}$ and \ $Q_{kR}^{a^{\prime }}$, where $a$
and $a^{\prime }\ $are indices of the corresponding metacolor subgroups $%
SO(3)_{L}$ ($a=1,2,3$) and $SO(3)_{R}$\ ($a^{\prime }$ ${=1,2,3}$),
respectively. They are generically anomaly-free and provide the minimal
three-preon configurations for composite quarks and leptons. Due to the
chiral metacolor, there are two types of composites at large distances being
composed individually from the left-right and left-handed preons,
respectively.

(3) Obviously, the preon condensate $\left\langle \overline{P}%
_{L}Q_{R}\right\rangle $ which could cause the metacolor scale $\Lambda
_{MC} $ order masses for composites is principally impossible in the
left-right metacolor model taken. This may be generally considered as a
necessary but not yet a sufficient condition for masslessness of composites.
The genuine massless fermion composites are presumably only those which
preserve the chiral $SU(K)_{L}\times SU(K)_{R}$ symmetry of preons at large
distances that is controlled by the 't Hooft's anomaly matching condition 
\cite{th}.

(4) The strengthening of this condition in a way that the massless fermion
composites, both left-handed and right-handed, are required to complete a
single representation of the $SU(K)_{MF}$ rather than some set of its
representations, allows to fix the number of basic metaflavors $K$.
Particularly, just eight left-handed and eight right-handed preons and their
composites preserving the global chiral symmetry $SU(8)_{L}\times SU(8)_{R}$
are turned out to uniquely identify the local metaflavor symmetry $%
SU(8)_{MF} $ as the grand unified symmetry of preons at small distances.

(5) An appropriate violation or the starting $L$-$R$ symmetry at large
distances breaks then this vectorlike $SU(8)_{MF}$ symmetry for preons down
to the chiral $SU(5)\times SU(3)_{F}$ symmetry for composites that contains
the conventional $SU(5)$ GUT with an extra local family symmetry $SU(3)_{F}$
and three standard families of composite quarks and leptons.

We can see now that in order to adapt the constrained gauge bosons to this
scenario one has to generate the local vectorlike metaflavor symmetry $%
SU(8)_{MF}$ together with the local chiral metacolor symmetry $%
SO(3)_{L}\times SO(3)_{R}$ binding separately the left-handed and
right-handed preons. This means that, while the metaflavor part in the above
preon Lagrangian (\ref{st}) is left intact, its metacolor part has to be
properly changed. Namely, it should now include separately both left-handed
and right-handed preons, $P_{kL}^{a}$ and $Q_{kR}^{a^{\prime }}$, being
assigned to the different metacolor groups $SO(3)_{L}$ and $SO(3)_{R}$
though to the same metaflavor symmetry $SU(8)_{MF}$. Also, there are the two
metacolor currents, $J_{\mu }^{a}$ and $J_{\mu }^{a^{\prime }}$ of $%
SO(3)_{L} $ and $SO(3)_{R}$ with the corresponding gauge bosons $A_{\mu
}^{a} $ and $A_{\mu }^{a^{\prime }}$, as well as the two constraining
Lagrange multiplier terms with the $\mathrm{\lambda }_{MC}$ and $\mathrm{%
\lambda }_{MC}^{\prime } $ fields in the modified Lagrangian. It is clear
that this Lagrangian is emerged in turn after integration out of the
corresponding massive constituent fermion fields $\psi _{k}^{a}$ and $\psi
_{k}^{a^{\prime }}$ which, in contrast to chiral preons, are vectorlike ($%
\psi _{k}^{a}\equiv (\psi _{L,R})_{k}^{a}$ , $\psi _{k}^{a^{\prime }}\equiv
(\psi _{L,R})_{k}^{a^{\prime }}$) with respect to their own symmetry group, $%
SO(3)_{L}$ or $SO(3)_{R}$, respectively. It is worth noting that these
constituents will produce themselves the composite states due metacolor
forces they possess. However, they will definitely appear very heavy (of the
order of the metacolor scale taken) and might only influence physics at high
energies comparable with the grand unification scale or so.

We have thus briefly demonstrated one particular way of unification of the
constrained composite gauge bosons with composite quarks and leptons that
could appear due to the effective preon Lagrangian (\ref{st}) and its
extensions. There are presumably some other ways as well.

\section{Conclusion}

We have considered the class of composite models for gauge fields, which are
emerged from the prototype fermion model taken in the current-current
interaction form. In contrast to the conventional approach where such
interactions leads generally to the unstable vector field potential, in our
model with the properly constrained fermion currents the composite gauge
bosons come out to be only restricted by the nonlinear covariant constraint
of type $A_{\mu }^{2}=M^{2}$. They appear naturally massless as the NG modes
of SLIV, so that the global internal symmetry $G$ in the model turns into
the local symmetry $G_{loc}$, while the vector field constraint reveals
itself as the gauge fixing condition\footnote{%
Generally, apart from the current-current interaction terms, there are many
other fermion bilinears and their interactions that could be included in our
prototype Lagrangian (\ref{fF}). The above "bosonization" procedure could
then be generalized so as to introduce a new auxiliary field (scalar, vector
or tensor one) for each bilinear that eventually would lead to an effective
action for a set of interacting auxiliary fields \cite{kraus}. In contrast
to the above massless $A_{\mu }$ fields induced by the current bilinears,
they are not constrained and, presumably, acquire large masses. On general
grounds, these masses have to be of the cutoff order and, therefore, all
such states can be neglected at low energies.}.

Whereas the field-current identity for the massive vector fields, underlying
the conventional argumentation, has a long story dating back to the paper of
Kroll, Lee and Zumino \cite{klz}, such an identity for massless vector
fields seems hardly possible unless they are properly constrained. The
crucial point, as we could see, happens to be the relation (\ref{rel})
between the standard matter currents and extra classical current introduced
by Dirac \cite{dir}. Actually, one now has in a sense the current-current
identity rather than field-current one.

We can go a bit further and wonder why basically the proposed constraint
pattern for vector fields could make the whole physical field system
involved to adjust itself in a gauge invariant way. The possible answer
seems to be that the only theories compatible with the nonlinear vector
field constraints taken are the gauge invariant ones, as has been generally
argued in \cite{cj} (see also \cite{cfn2, c3}). Indeed, once the SLIV
constraint (\ref{cd}) or \ref{aa}) is imposed, it is therefore not possible
to satisfy another  supplementary condition since this would superfluously
restrict the number of degrees of freedom for the vector field. To avoid
this, its equation of motion should be automatically divergenceless since
otherwise one would have one more condition. However, such an equation of
motion is only possible in the gauge invariant theory. Actually, gauge
invariance in theories considered appears in essence as a response of an
interacting field system\textit{\ }to putting the covariant constraint (\ref%
{cd}, \ref{aa}) on its dynamics, provided that we allow parameters in the
corresponding Lagrangian density to be adjusted so as to ensure
self-consistency without losing too many degrees of freedom. Otherwise, a
given field system could get unphysical in a sense that a superfluous
reduction in the number of degrees of freedom would make it impossible to
set the required initial conditions in the appropriate Cauchy problem.
Namely, it would be impossible to specify arbitrarily the initial values of
the vector and other field components involved, as well as the initial
values of the momenta conjugated to them. Furthermore, in quantum theory, to
choose self-consistent equal-time commutation relations would also become
impossible \cite{ogi3}. So, the nonlinear SLIV condition (\ref{cd}, \ref{aa}%
), due to which true vacuum in the theory is chosen and massless gauge
fields are generated, may provide a dynamical setting for all underlying
internal symmetries involved through such an emergence conjecture.

One can see that the gauge theory framework, be it taken from the outset or
emerged, makes in turn SLIV to be physically unobservable both in Abelian
and non-Abelian symmetry case that is also favored by the supersymmetric
SLIV version. We referred to it above as the inactive SLIV in contrast to
the active SLIV case where physical Lorentz violation could effectively
occur. From the present standpoint, the only way for an inactive SLIV to be
activated would be if emergent gauge symmetries presented above were
slightly broken at small distances being presumably controlled by quantum
gravity. One might even think that quantum gravity could in principle hinder
the setting of the required initial conditions in the appropriate Cauchy
problem, thus admitting a superfluous restriction of vector fields in terms
of some high-order operators which occur at the Planck scale order
distances. This is just the range of distances where the composite gauge
fields, as well as the composite quarks and leptons, should presumably
emerge in order to be properly adapted to the grand unification landscape.
So, some trace of the gauge symmetry breaking and therefore physical Lorentz
violation might accompany their composition processes. We may return to this
special issue elsewhere.

\section*{ Acknowledgments}

I would like to thank Riccardo Barbieri, Harald Fritzsch, Colin Froggatt,
Oleg Kancheli, Rabi Mohapatra and Holger Nielsen for stimulating discussions
on many aspects of grand unified theories, family replications and composite
nature of quark-lepton species and gauge fields.


\begin{thebibliography}{99}
\bibitem{NJL} Y.~Nambu, G.~Jona-Lasinio, Phys. Rev. 122 (1961) 345;

J. Goldstone, Nuovo Cimento 19 (1961) 154.

\bibitem{bjorken} J.D.~Bjorken, Ann. Phys. (N.Y.) \ 24 (1963) 174.

\bibitem{ph} P.R. Phillips, Phys. Rev. 146 (1966) 966.

\bibitem{eg} T.~Eguchi, Phys.Rev. D 14\textbf{\ }(1976)\textbf{\ }2755;

H. Terazava, Y. Chikashige, K. Akama, Phys. Rev. D 15 (1977) 480.

\bibitem{suz} M. Suzuki, Phys. Rev. D 37 (1988) 210; Phys. Rev. D 82 (2010)
045026.

\bibitem{kraus} Per Kraus, E.T. Tomboulis, Phys. Rev. D 66 (2002) 045015.

\bibitem{ks} V.A. Kostelecky, S. Samuel, Phys. Rev. D 39 (1989) 683.

\bibitem{vru} R.~Bluhm, N.L.~Gagne, R.~Potting, A.~Vrublevskis, Phys.\ Rev.\
D 77 (2008) 125007.

\bibitem{dr} P.A.M. Dirac, \textit{Lectures on Quantum Mechanics} (Yeshiva
University, New York, 1964).

\bibitem{aki} A. Hashimoto, JHEP 0808 (2008) 040.

\bibitem{c} J.L. Chkareuli, Phys. Lett. B 721 (2013) 146; Phys. Rev. D 90
(2014) 065015.

\bibitem{dir} P.A.M. Dirac, Proc. Roy. Soc. A 209 (1951) 291. \ \ \ \ \ \ \
\ \ \ \ \ \ \ \ \ \ \ \ \ \ \ \ \ \ \ \ \ \ \ \ \ \ \ \ \ \ \ \ \ \ \ \ \ \
\ \ \ \ \ \ \ \ \ \ 

\bibitem{nambu} Y. Nambu, Progr. Theor. Phys. Suppl. Extra 190 (1968).

\bibitem{az} A.T. Azatov, J.L. Chkareuli, Phys. Rev. D 73 (2006) 065026.

\bibitem{ks1} R. Bluhm, S.-H. Fung, V.A. Kostelecky, Phys. Rev. D 77 (2008)
065020.

\bibitem{jej} J.L. Chkareuli, J.G. Jejelava, Phys. Lett. B 659 (2008) 754.

\bibitem{cj} J.L. Chkareuli, C.D. Froggatt, J.G. Jejelava, H.B. Nielsen,
Nucl. Phys. B 796\textbf{\ }(2008) 211.

\bibitem{wiki} J.L. Chkareuli, C.D. Froggatt, H.B. Nielsen, Nuclear Physics
B 821 (2009) 65.

\bibitem{gra} J.L. Chkareuli, J.G. Jejelava, G. Tatishvili, Phys. Lett. B
696 (2011) 124.

\bibitem{urr} O.J. Franca, R. Montemayor, L.F. Urrutia, Phys. Rev. D 85
(2012) 085008.

\bibitem{esc} C.\thinspace A. Escobar, L.\thinspace F. Urrutia, Phys. Rev.
D~92 (2015) 025013; Phys. Rev. D~92 (2015) 025042.

\bibitem{GL} S. Weinberg, \textit{The Quantum Theory of Fields}, vol.2
(Cambridge University Press, Cambridge 2000).

\bibitem{ll} J.L. Chkareuli, Phys. Lett. B 810 (2020) 135625.

\bibitem{par} J.L. Chkareuli, Z. Kepuladze, G. Tatishvili, Eur. Phys. J. C
55 (2008) 309;

J.L. Chkareuli, Z. Kepuladze, Eur. Phys. J. C 72 (2012) 1954.

\bibitem{cfn} J.L.~Chkareuli, C.D.~Froggatt, H.B.~Nielsen,
Phys.~Rev.~Lett.~87 (2001)\ 091601; Nucl.~Phys.~B \ 609 (2001) 46.

\bibitem{pr} R.N. Mohapatra, \textit{Unification and Supersymmetry}
(Springer-Verlag, New-York, 2003).

\bibitem{th} G. 't Hooft, in \textit{Recent Developments in Gauge Theories},
edited by G.'t Hooft et al (Plenum, New-York, 1980).

\bibitem{su} J.L. Chkareuli, Nucl.~Phys.~B 941 (2019) 425.

\bibitem{cfn2} J.L. Chkareuli, C.D. Froggatt, H.B. Nielsen, Nucl. Phys. B
848 (2011) 498.

\bibitem{c3} J.L. Chkareuli, On Emergent Gauge and Gravity Theories, in 
\textit{Low dimensional physics and gauge principles, }pp 80-92 (World
Scientific, Singapore, 2011); arXiv:1206.1368 [hep-th].

\bibitem{klz} N.M. Kroll, T.D. Lee, B. Zumino, Phys. Rev. 157 (1967) 1376.

\bibitem{ogi3} V.I.~Ogievetsky, I.V. Polubarinov, Ann. Phys. (N.Y.) 25%
\textbf{\ }(1963) 358; Nucl. Phys. 76 (1966) 677.
\end{thebibliography}
\end{document}